\documentclass[twocolumn,aps, prb, superscriptaddress,floatfix,nobibnotes]{revtex4-1}
\usepackage{graphicx} 
\usepackage{amssymb} %fancy symbols package
\usepackage{xcolor} %fancy color package to make a pretty article
\usepackage{amsmath} %fancy math package
\usepackage{hyperref} %fancy hyperlink package
\usepackage{epstopdf} %package to deal with eps files
\usepackage[nointegrals]{wasysym} %fancier symbols package, must be called AFTER amsmath

\graphicspath{ {Figures/} } %looks for figures in the Figures folder

\begin{document}
	
\title{Effects of chemical disorder in the itinerant antiferromagnet Ti$_{1-x}$V$_x$Au}
\author{C.-L. Huang}
\email[]{clh@rice.edu}
\affiliation{Department of Physics and Astronomy, Rice University, Houston, TX 77005}

\author{J.M. Santiago}
\affiliation{Department of Physics and Astronomy, Rice University, Houston, TX 77005}
\author{E. Svanidze}
\affiliation{Department of Physics and Astronomy, Rice University, Houston, TX 77005}
\author{T. Besara}
%\altaffiliation[Current address: ]
%{Department of Physics, Missouri State University, Springfield, MO}
\affiliation{National High Magnetic Field Laboratory, Florida State University, Tallahassee, Florida, 32306}
\author{T. Siegrist}
\affiliation{National High Magnetic Field Laboratory, Florida State University, Tallahassee, Florida, 32306}
\author{E. Morosan}
\affiliation{Department of Physics and Astronomy, Rice University, Houston, TX 77005}

\date{\today}

\begin{abstract}
The fragile nature of itinerant magnetism can be exploited using non-thermal parameters to study quantum criticality. The recently discovered quantum critical point (QCP) in the Sc-doped (hole-like doping) itinerant antiferromagnet TiAu (Ti$_{1-x}$Sc$_{x}$Au) raised questions about the effects of the crystal and electronic structures on the overall magnetic behavior. In this study, doping with V (electron-like doping) in Ti$_{1-x}$V$_{x}$Au introduces chemical disorder which suppresses antiferromagnetic order from $T_{\rm N} =$ 36~K for $x = 0$ down to 10 K for $x =$ 0.15, whereupon a solubility limit is reached. Signatures of non-Fermi-liquid behavior are observed in transport and specific heat measurements similar to Ti$_{1-x}$Sc$_{x}$Au, even though Ti$_{1-x}$V$_{x}$Au is far from a QCP for the accessible compositions $x \leq 0.15$. 
\end{abstract}

\maketitle %makes a title page

\section{Introduction} \label{intro}
The recently discovered itinerant antiferromagnetic metal TiAu\cite{svanidze_itinerant_2015} is only the third itinerant magnet (IM) with no magnetic elements, after the two ferromagnets ZrZn$_2$ \cite{Uhlarz2004} and Sc$_3$In\cite{grewe_anomalous_1989,svanidze_non-fermi_2015} discovered several decades ago. The itinerant antiferromagnet (IAFM) TiAu with a N\'eel temperature $T_{\rm N} =$ 36 K provides a new playground to perturb the fragile magnetism with non-thermal tuning parameters such as doping \cite{svanidze_quantum_2017} and pressure \cite{wolowiec_pressure_2017}. Such studies afford the tuning of long range magnetic order towards a quantum critical point (QCP), i.e., a second order phase transition at zero temperature driven by quantum fluctuations, where non-Fermi-liquid (NFL) behaviors were observed \cite{hertz_quantum_1976,millis_effect_1993,lohneysen_fermi-liquid_2007,brando_metallic_2016}. 
	
	The application of pressure $p$ initially results in a small increase of $T_{\rm N}$ in TiAu, akin to what had been observed in the itinerant ferromagnet Sc$_{3}$In \cite{grewe_anomalous_1989}. Subsequent pressure increase in TiAu suppresses $T_{\rm N}$ to roughly 22 K around $p =$ 27 GPa, and a pressure-induced QCP has been estimated to emerge at a critical pressure of $p = 45$ GPa \cite{wolowiec_pressure_2017}. Chemical doping can mimic positive or negative pressure, depending on the relative size of the dopant compared to the host atoms, introduce additional carriers (charge doping), or induce disorder. This provides more versatility than applied pressure alone in tuning the magnetic ground state. Upon doping with Sc in Ti$_{1-x}$Sc$_{x}$Au, $T_{\rm N}$ was suppressed to a QCP around $x_{c}$ = 0.13 where NFL behavior was observed in temperature-dependent electrical resistivity $\rho(T)$ and specific heat $C_{p}(T)$ measurements \cite{svanidze_quantum_2017}. In this case, the disorder effects were deemed minimal, as was chemical pressure, given the similarity in the size of the Ti and Sc ions\cite{shannon_revised_1976,greenwood_chemistry_2012}. 
	
	In this study, we turn to doping with V in Ti$_{1-x}$V$_{x}$Au. In addition, as in the Sc case, V is chosen to minimize chemical pressure, given the metallic radius r[V] = 1.34~\AA with respect to r[Ti] = 1.47~\AA\cite{shannon_revised_1976,greenwood_chemistry_2012}. In contrast to the Sc doping, increasing disorder brought about by V doping is evident in both x-ray and resistivity data. $T_{\rm N}$ is suppressed down to $\sim$ 10 K with $x$ = 0.15, where the solubility limit of V in TiAu is reached. Signatures of NFL behavior are observed in $\rho(T)$ and $C_{p}(T)$. Comparisons are drawn between doping with Sc and with V in TiAu.

	\section{Experimental Methods} \label{methods}
	Polycrystalline samples of Ti$_{1-x}$V$_{x}$Au were arc melted in stoichiometric ratios of Ti (99.9\% pieces, Alfa Aesar), V (99.7\% pieces, Alfa Aesar), and Au (99.999\% splatter, Materion) under an inert Ar atmosphere. The arc melted button was remelted several times to ensure homogeneity, and mass losses were $\leq$ 0.3\% for all samples. All data are presented with the nominal doping concentration $x$. The crystal structure was determined via x-ray powder diffraction on the polished cross sections of arc melted samples, as these samples could not be ground into a powder due to extreme hardness \cite{svanidze_itinerant_2015,svanidze_high_2016}. The x-ray diffraction was performed on a custom 4-circle Huber diffractometer with graphite monochromator and analyzer in a non-dispersive geometry, along with a Rigaku rotating CuK$\alpha$ anode source.
			
	The DC magnetic susceptibility was measured in a Quantum Design (QD) Magnetic Property Measurement System from 1.8 to 300 K in a field of $\mu_0 H$ = 0.01 T. Four-probe electrical resistivity and specific heat measurements were performed in a QD Physical Property Measurement System with a 3He insert down to 0.35 K in zero field. The resistivity was measured using an AC excitation current $\leq$ 1.3 mA and with frequency multiples of $f$ = 17.77 Hz up to 604.18 Hz. The specific heat using an adiabatic relaxation method of the sample was determined via background subtraction of the Apiezon N grease.
	
	\section{Results and Analysis} \label{results}
\begin{figure}[h]
		\centering
		\includegraphics[width=\columnwidth]{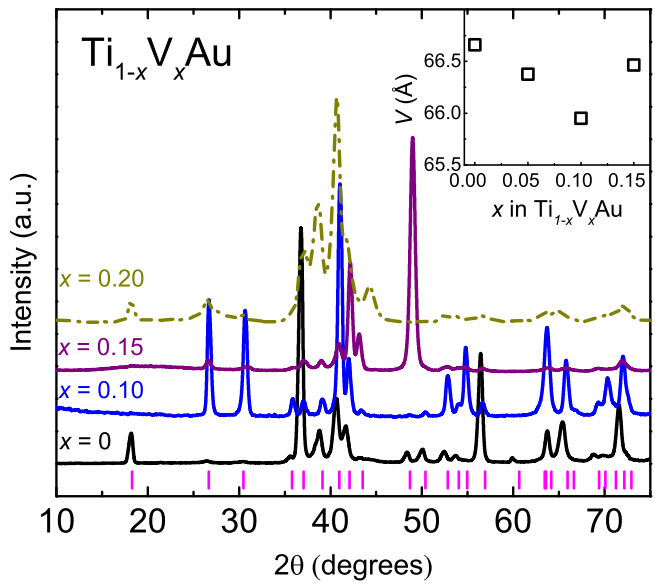}
		\caption{A subset of Ti$_{1-x}$V$_{x}$Au x-ray powder diffraction patterns for $0\leq x \leq0.20$. The calculated peak positions for $Pmma$ TiAu are marked by vertical lines (bottom). Inset shows an evolution of the unit cell volume with increasing doping concentration $x$ up to the solubility limit at $x$~=~0.15.}\label{fig:xrd}
	\end{figure}

	%XRD results
	Figure~\ref{fig:xrd} shows x-ray diffraction patterns of Ti$_{1-x}$V$_{x}$Au with $x = 0, 0.1, 0.15,$ and 0.20. Due to the hardness of the samples, strong preferred orientation in the surface enhances the intensities of some Bragg reflections. From a structural point of view, doping with V up to $x$ = 0.15 in Ti$_{1-x}$V$_{x}$Au does not appear to induce a significant amount of chemical pressure, with a unit cell volume change of $\sim$ 1\% (inset of Fig.~\ref{fig:xrd}), roughly $2/3$ of that induced by Sc doping in Ti$_{1-x}$Sc$_{x}$Au for comparable $x$ \cite{svanidze_quantum_2017}. However, the increasingly broad peaks as more V dopes into the orthorhombic $Pmma$ lattice of TiAu\cite{donkersloot_martensitic_1970,massalski_binary_1990} indicate a solubility limit above $x$ = 0.15. A qualitative change in the x-ray patterns is apparent between $x \leq$ 0.15 (purple curve in Fig.~\ref{fig:xrd}) and $x$ = 0.20 (dashed line in Fig.~\ref{fig:xrd}), with most peaks still present but much broader in the latter pattern. A number of factors may be at play, such as internal stress and strain due to the increasing amount of V, or disorder induced by V doping. %It should be noted that the lower compositions show all equivalent peaks for the TiAu $Pmma$ structure, albeit with variable peak intensities. This can be attributed to variable crystallite orientation in the surface of the samples used for these measurements. 
The situation was very different in Ti$_{1-x}$Sc$_{x}$Au, where the diffraction peaks did not appear to substantially change in width up to $x$ = 0.25 (see Supplementary Material of Ref.~\onlinecite{svanidze_quantum_2017}). These structural changes with Sc and V doping point to a %much weaker $x$ dependence in the latter case, and also a 
greater amount of disorder in the latter which likely dominates the physical properties as described below.
	
\begin{figure}[h!]
	
		\centering
		\includegraphics[width=\columnwidth]{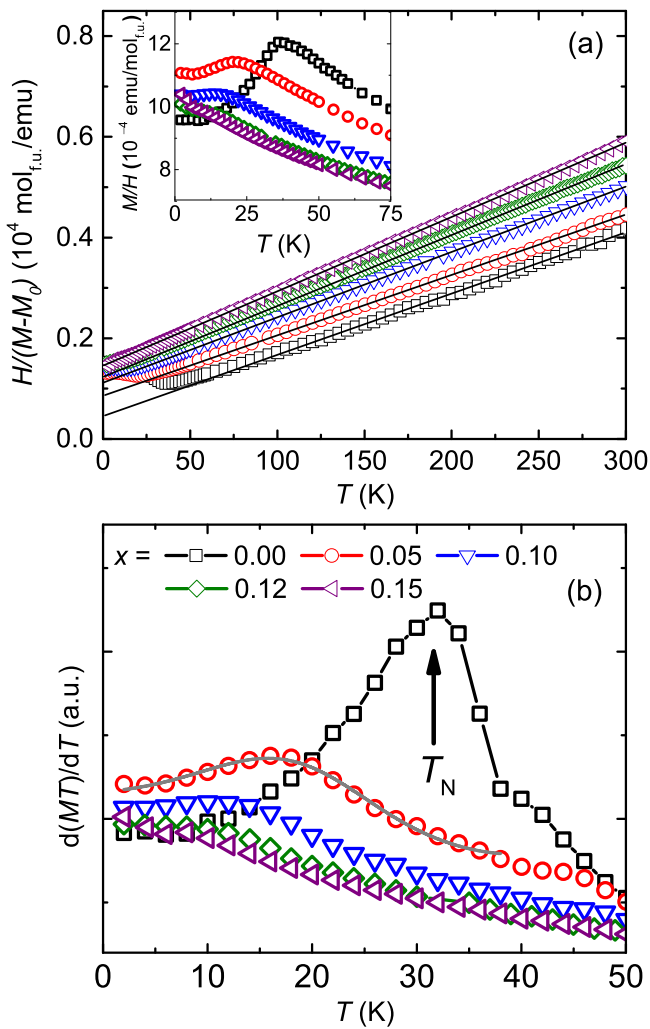}
		\caption{(a) The temperature-dependent inverse magnetic susceptibility of Ti$_{1-x}$V$_{x}$Au for $0\leq x \leq0.15$. Curie-Weiss-\textit{like} fits for each concentration $x$ are shown in solid gray lines. Inset shows the antiferromagnetic cusps in the temperature-dependent $M/H$ around $T_{\rm N}$. Temperature independent contribution $M_0$ has been subtracted for all samples. (b) The determination of $T_{\rm N}$ from d$(MT)$/d$T$, with an example of a peak fit shown in gray for $x$ = 0.05.}\label{fig:MT}
	\end{figure}

	%MT results
	The magnetic susceptibility $M/H$ of Ti$_{1-x}$V$_{x}$Au shows an antiferromagnetic cusp being suppressed with increasing $x$ (inset of Fig.~\ref{fig:MT}(a)). A tiny amount of substitution with $x$ = 0.05 substantively suppresses $T_{\rm N}$ by $\sim$ 30$\%$ (from 36 K in pure TiAu), and the cusp itself broadens as the transition shifts to lower temperatures. By $x$ = 0.15, the cusp nearly vanishes in a large Pauli susceptibility-like background. Antiferromagnetic order is determined at a peak position in a d$(MT)$/d$T$ plot as shown in Fig.~\ref{fig:MT}(b) \cite{fisher_relation_1962}. The self-consistent renormalization (SCR) theory for spin fluctuations\cite{takahashi_origin_1986,moriya_spin_2012} describes the Curie-Weiss-\textit{like} temperature-dependence of the inverse magnetic susceptibility $H/(M-M_{0})$ (Fig.~\ref{fig:MT}(a)). Analogous to the effective moment $\mu_{eff}$ for local moment magnets, the paramagnetic moment $\mu_{PM}$ of the itinerant system can be obtained from a linear fit of $H/(M-M_{0})$ for $T > T_{\rm N}$, shown as solid lines in Fig.~\ref{fig:MT}(a). The value of $\mu_{PM}$ decreases only slightly across the series, and the Weiss-\textit{like} temperature $T^{*}$, analogous to the local moment Curie-Weiss temperature $\theta_{\rm W}$, appears to follow a similar trend as $T_{\rm N}$. These values are summarized in the $T - x$ phase diagram below. 	
	
	The suppression of antiferromagnetic order with V doping is found to be drastically different from Sc doping in the IAFM TiAu. In Ti$_{1-x}$Sc$_{x}$Au, $T_{\rm N}$ is continually suppressed to 0 K at a critical concentration of $x_{c}$~=~0.13, and the $T_{\rm N}(x)$ dependence is consistent with the SCR theory, reflecting the role of spin fluctuations  \cite{hasegawa_effect_1974}. By contrast, the antiferromagnetic order persists at $x$ = 0.15 in Ti$_{1-x}$V$_{x}$Au, where $T_{\rm N} \sim$ 10~K. Both Sc and V doping in TiAu exhibit Curie-Weiss-\textit{like} behavior. However, $\mu_{PM} \sim$ 0.8 $\mu_{\rm B}$ f.u.$^{-1}$ for pure TiAu barely changes with increasing Sc substitution, even upon approaching the QCP at $x_{c}$  \cite{svanidze_quantum_2017}, while $\mu_{PM}$ decreases by $\sim$10\% at $x =$ 0.15 for the V-doped case. This suggests that, although Sc and V generate the similar amount of chemical pressure  in TiAu, they play different roles for the suppression of magnetic order. 
		
	%CpT results
	\begin{figure}[h]
		\centering
		\includegraphics[width=\columnwidth]{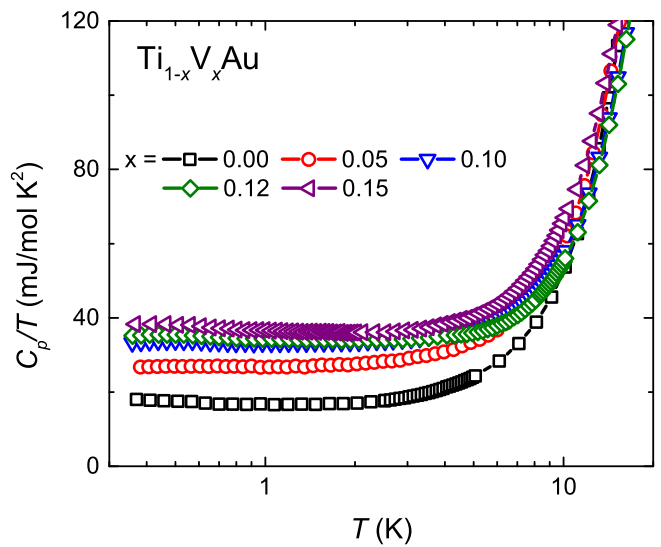}
		\caption{Plot of $C_{p}/T$ vs. $\log T$ highlights the low-temperature specific heat behavior of Ti$_{1-x}$V$_{x}$Au for $0\leq x \leq0.15$.}\label{fig:CpT}
	\end{figure}
	
	In Ti$_{1-x}$Sc$_{x}$Au the proximity to a QCP was indicated by an increasingly divergent specific heat $C_{p}/T$ at low temperatures\cite{svanidze_quantum_2017}. Similar NFL behavior is found in Ti$_{1-x}$V$_{x}$Au. $C_{p}/T$ is nearly constant below 2~K for $x \leq$ 0.10), while $C_{p}/T$ begins to rise below 2 K for $x$ = 0.12, and shows weak logarithmic divergence for nearly a decade in temperature for $x$ = 0.15 below $T \leq$ 3 K, as shown in Fig.~\ref{fig:CpT}). Additionally, the electronic component of the specific heat $\gamma$ nearly doubles with $x$ = 0.15 compared to that of pure TiAu and the value is comparable to that determined in Ti$_{1-x}$Sc$_{x}$Au with $x_c =$ 0.12\cite{svanidze_quantum_2017}, despite the absence of a QCP in Ti$_{1-x}$V$_{x}$Au.
	
	%RT results
	\begin{figure}
		\centering
		\includegraphics[width=\columnwidth]{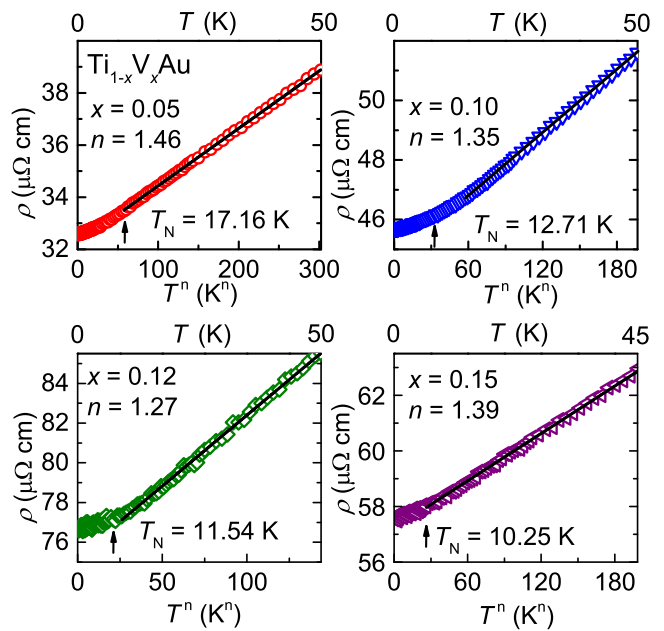}
		\caption{The temperature-dependent electrical resistivity of Ti$_{1-x}$V$_{x}$Au for concentrations $x$ = 0.05, 0.10, 0.12 and 0.15 plotted as a function of $T^{n}$. The straight lines demonstrates fits to the expression $\rho(T) = \rho_{0} + T^{n}$ in the paramagnetic region $T > T_{\rm N}$, where $T_{\rm N}$ is indicated by an arrow.}\label{fig:RT}
	\end{figure}
	
	With signatures of NFL behavior appearing in $C_{p}(T)$ at low temperatures, the electrical resistivity  was also investigated in Ti$_{1-x}$V$_{x}$Au to corroborate these findings. Fitting of the low-temperature resistivity to $\rho (T) = \rho_{0} + A T^{n}$, where $\rho_{0}$ is the residual resistivity, determines the strength of the temperature dependence $n$, as shown in Fig.~\ref{fig:RT}. Above $T_{\rm N}$, $n \approx 2$ for pure TiAu, consistent with a Fermi liquid picture. With introducing the V substitution, $n$ decreases to less than 1.5, a signature of NFL behavior. This is illustrated by the linear fits (black lines) of $\rho$ vs. $T^{n}$ in Fig.~\ref{fig:RT}.  The electrical transport parameters for the Ti$_{1-x}$V$_{x}$Au series are summarized in Table~\ref{Table1} along with data from Ref.~\onlinecite{svanidze_quantum_2017} on the analogous Sc-doped series. The residual resistivity ratio RRR = $\rho(300K)/\rho_0$ for Ti$_{1-x}$V$_{x}$Au generally decrease with increasing $x$, possibly because of increasing disorder. This is also reflected in the increasing residual resistivity values by one order of the magnitude, from $\rho_{0} =$ 7.2 $\mu\Omega$ cm for $x = 0$ to the maximum of $\sim$77 $\mu\Omega$ cm for $x = 0.12$. The temperature dependence of $\rho(T)$ is comparably weaker in Ti$_{1-x}$V$_{x}$Au than in the case of Sc doping, i.e., smaller $n$ values for the former. The resistivity difference, $\Delta\rho = \rho - \rho_0$ in the region where the relation $\Delta\rho \sim T^{n}$ holds, is much less than $\rho_{0}$, and the ratio of $\Delta\rho/\rho_0$ is smaller in the V doped series than in the Sc doped one, though both series have similar values of RRR \cite{svanidze_quantum_2017}. 
	
	\begin{table*}[ht]
		\caption{\label{Table1} Parameters determined from electrical resistivity measurements and exponents extrapolated from fitting the data to the function $\rho(T) = \rho_{0} + A T^{n}$ for Ti$_{1-x}$M$_{x}$Au where M = V in this study and M = Sc from Ref.~\onlinecite{svanidze_quantum_2017}.}
		\begin{center}
			%\scalebox{0.7}{
			\begin{tabular}{c|c|c|c|c|c|c}
			\hline \hline
			$x$ & \multicolumn{3}{c|}{\textbf{M = V}} & \multicolumn{3}{c}{\textbf{M = Sc}} \\ \cline{2-7}
			& RRR & $\rho_{0}$ ($\mu\Omega$cm) & $n$ ($T > T_{\rm N}$) & RRR & $\rho_{0}$ ($\mu\Omega$cm) & $n$ ($T > T_{\rm N}$) \\ \hline \hline
			0.00$^{a}$ & 6.1 & 7.2   & 1.78 & 6.1  & 7.2   & 1.78   \\ \hline
			0.05       & 2.5 & 32.59 & 1.47 & 1.7 & 15.23 & 1.88 \\ \hline
			0.10       & 2.0 & 45.69 & 1.36 & 2.1 & 29.65 & 1.76 \\ \hline
			0.12       & 1.8 & 76.70 & 1.27 & 2.0 & 36.92 & 1.23 \\ \hline
			0.15       & 1.7 & 57.94 & 1.39 & 1.8 & 43.63 & 1.24  \\ \hline \hline
			\multicolumn{4}{l}{\footnotesize $^{a}$ From Ref.~\onlinecite{svanidze_itinerant_2015}}	
			\end{tabular}
			\end{center}
			
	\end{table*}
	
	\section{Discussion} \label{end}	
	%The IAFM TiAu has been taken down several avenues in order to elucidate its rather rare magnetic properties. In the present study, V doping in TiAu partially suppresses the itinerant antiferromagnetic order to $T_{\rm N} \approx$ 10 K at $x$ = 0.15. 
	%Signatures of NFL behavior in Ti$_{1-x}$V$_{x}$Au are revealed in both the specific heat and electrical resistivity data, as summarized in Fig.~\ref{fig:phaseDiagram}. 
	The difference in metallic radii is larger between Ti and Sc than between Ti and V, resulting in a larger unit cell volume change in Ti$_{1-x}$Sc$_{x}$Au $\sim 2\%$ for $x$ up to 0.13, compared to only $\sim 1\%$ for comparable V doping. This volume effect could, at least partially, explain why the magnetism is suppressed faster with Sc than V doping. Electronic structure calculations revealed a large peak in the density of states just above the Fermi surface $E_{\rm F}$ in pure TiAu \cite{goh_mechanism_2016}. From a rigid band perspective, Sc doping is treated as introducing hole carrier that shifts the peak away from $E_{\rm F}$, while V doping is treated as electron doping that shifts the peak to lower energy, crossing $E_{\rm F}$. An asymmetric phase diagram was therefore predicted, where electron (V) doping actually enhanced the ordered moment slightly before suppressing the magnetic order, albeit less effectively than electron (Sc) doping \cite{goh_competing_2017}. No such enhancement was observed in Ti$_{1-x}$V$_{x}$Au, and on the contrary the ordered moment is reduced as increasing the V concentration, as shown in the inset of Fig.~\ref{fig:MT}(a). Such discrepancy between theory and experiment might be explained by the fact that the amount of disorder in Ti$_{1-x}$V$_{x}$Au is non-negligible, while the theoretical calculation by Goh $et$ $al.$ completely ignores disorder effect \cite{goh_competing_2017}. In addition, the role of electron or hole doping is ambiguous for intermetallic systems where the type of chemical bonding is mainly metallic bonding.  

	\begin{figure}[h]
		\centering
		\includegraphics[width=\columnwidth]{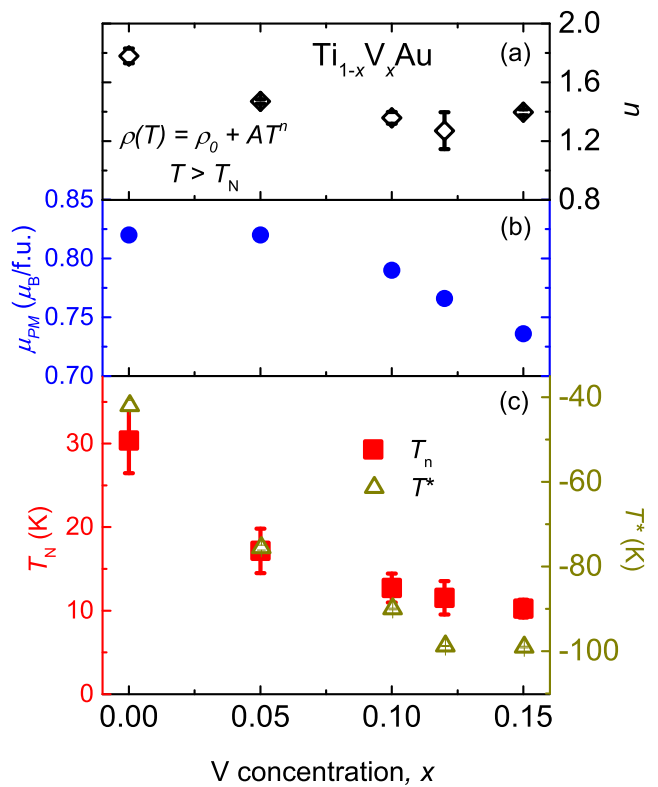}
		\caption{The $T - x$ phase diagram of Ti$_{1-x}$V$_{x}$Au for $0\leq x \leq0.15$. The evolution of the Weiss-\textit{like} temperature $T^{*}$, N\'eel temperature $T_{\rm N}$, paramagnetic moment $\mu_{PM}$, and temperature dependence of the resistivity $n$ are tracked across the sample series.}\label{fig:phaseDiagram}
	\end{figure} 
		
	Both the x-ray diffraction patterns and $\rho(T)$ measurements demonstrate signs of disorder in Ti$_{1-x}$V$_{x}$Au, as evidenced by increasingly broad diffraction peaks and large $\rho_{0}$ values. Also, $\rho_{0}$ exceeds the change in resistivity $\Delta \rho$ in the region of $\rho(T)$ linear in $T^{n}$, another indication of strong disorder scattering. Finally, the Ti$_{1-x}$V$_{x}$Au series is terminated at the doping limit of $x >$ 0.15, most probably due to strong disorder that precludes phase formation. But could this disorder be helping to suppress the antiferromagnetic order? A classical critical point, i.e., a phase transition at non-zero temperature driven by thermal fluctuations, is stable against quenched disorder, while the QCP in an itinerant magnet is strongly affected by the presence of disorder \cite{vojta_rare_2006}. In the latter case, theoretical investigations have shown that disorder can smear a phase transition or even prevent the formation of long range order  \cite{vojta_rare_2006,narayanan_critical_1999,dobrosavljevic_absence_2005}. When disorder is strong, spatial fluctuations of disorder form rare regions devoid of impurities, and these rare regions either develop a finite region around the QCP called the Griffiths phase giving rise to the power-law temperature dependence of thermal dynamic observables (for example, Ni$_{1-x}$V$_x$ \cite{Kassis2010}), or form a locally ordered cluster spin glass state (for example, Sr$_{1-x}$Ca$_x$RuO$_3$ \cite{Demko2012,Fuchs2015}). These collective behaviors due to strong disorder have so far only been observed in itinerant ferromagnet systems with magnetic constituents, and whether these behaviors also exist in the IAFM TiAu without magnetic constituents is unknown. Our magnetic susceptibility and specific heat data in Ti$_{1-x}$V$_{x}$Au do not show a continuous power-law scaling between $x = 0 - 0.15$, an evidence of the Griffiths phase, and AC susceptibility measurements (not shown) do not show frequency dependence up to 10,000 Hz. Therefore, how the magnetic properties of V-doped TiAu manifest at low temperatures might not be readily explained at this stage, hinting at an attractive yet elusive intertwined structure-property relationship that needs to be explored further in this itinerant system. Different growth methods are indispensable to diminish the effects of strain, disorder, and phase separation that could not be avoided in the arc melted samples in current study.

	\begin{acknowledgments} \label{ack}
	The work at Rice University was funded by the NSF DMR-1506704 (J.M.S., E. S., C.L.H. and E.M.). T.B. and T.S. are supported by award DE-SC0008832 from the Materials Sciences and Engineering Division in the U.S. Department of Energy's Office of Basic Energy Sciences and the National High Magnetic Field Laboratory through the NSF Cooperative Agreement No. DMR-1157490 and the State of Florida.
	\end{acknowledgments}	
			
\bibliographystyle{unsrt}
\bibliography{tivauRefs}

\end{document}